\documentclass[prl,twocolumn,superscriptaddress]{revtex4}

\usepackage{amsfonts,xfrac}
\usepackage{amsmath}
\usepackage{amssymb}
\usepackage{graphicx}
\usepackage{dcolumn}
\usepackage{dsfont}
\usepackage{times}
\usepackage{epsfig,color}
\usepackage[]{units}

\def\epsilon{\varepsilon}
    
\def\beq{\begin{equation}}
\def\eeq{\end{equation}}

\begin{document}

\title{Damped-Driven Granular Crystals: An Ideal Playground for Dark Breathers and Multibreathers}
\author{C. Chong$^*$}
\affiliation{Department of Mathematics and Statistics, University of Massachusetts,
Amherst MA 01003-4515, USA}
\author{F. Li$^*$}
\affiliation{Graduate Aerospace Laboratories (GALCIT) California Institute of Technology, Pasadena, CA 91125, USA}
\author{J. Yang}
\affiliation{Aeronautics and Astronautics, University of Washington,
Seattle, WA 98195-2400}
\author{M.O. Williams}
\affiliation{Department of Chemical and Biological Engineering
and PACM, Princeton University, Princeton, NJ, 08544, USA}
\author{I.G. Kevrekidis}
\affiliation{Department of Chemical and Biological Engineering
and PACM, Princeton University, Princeton, NJ, 08544, USA}
\author{P. G. Kevrekidis}
\affiliation{Department of Mathematics and Statistics, University of Massachusetts,
Amherst MA 01003-4515, USA}
\author{C. Daraio}
\affiliation{Department of Mechanical and Process Engineering (D-MAVT), 
Swiss Federal Institute of Technology (ETH),
8092
Zurich, Switzerland}

\date{\today}


\begin{abstract}
By applying an out-of-phase actuation at the boundaries of a uniform chain of granular particles, we demonstrate experimentally
that time-periodic and spatially localized structures with a nonzero background (so-called dark breathers) emerge for a wide range of parameter values and initial conditions. Importantly, the number of ensuing breathers 
within the  multibreather pattern produced 
can be ``dialed in'' by varying the frequency or amplitude of the actuation.
The values of the frequency (resp. amplitude) where the transition between 
different multibreather states occurs are predicted accurately
by the proposed theoretical model, which is  numerically shown to support exact dark breather solutions.  
The existence, linear stability, and bifurcation structure of the theoretical dark breathers are also studied in detail.
Moreover, the distributed sensing technologies developed herein 
enable a detailed space-time probing of the system and a 
systematic favorable comparison
between theory, computation and experiments.
\end{abstract}

\maketitle


{\it Introduction.} The study of discrete breathers has been a topic of intense
theoretical and experimental interest during the 25 years
since their theoretical inception, as has been recently
summarized e.g. in~\cite{Flach2007}. Among the broad and diverse list 
of fields
where such time-periodic, exponentially localized in space
structures have been of interest, we mention for instance
 optical waveguide
arrays or photorefractive crystals~\cite{moti},
micromechanical cantilever arrays~\cite{sievers},
Josephson-junction ladders~\cite{alex},
layered antiferromagnetic crystals~\cite{lars3},
halide-bridged transition metal complexes~\cite{swanson},
dynamical models of the DNA double strand \cite{Peybi}
and Bose-Einstein condensates in optical lattices~\cite{Morsch}.
Remarkably, however, most of these investigations have been
restricted to the context of bright such states, namely ones
supported on a {\it vanishing} background. Dark breathers, i.e.,
breather states on top of a {\textit{non-vanishing}} background
have been far less widely studied. Their recent robust 
realization in contexts such as surface water waves~\cite{Chabchoub}, 
or Bose-Einstein condensates~\cite{weller} (see also~\cite{djf} 
for a recent review), ferromagnetic film strips~\cite{carr}
or optical waveguide arrays (see e.g. for a recent example~\cite{kanshu}
and references therein)
has received considerable attention~\footnote{It is important
to also note in passing that these dark solitonic states identified
experimentally are so-called envelope solitons and not dark breathers
in the spirit proposed theoretically 
in nonlinear oscillator systems~\cite{Rey04,Alvarez02}.}.

\begin{figure*}[t]
\centerline{
\epsfig{file=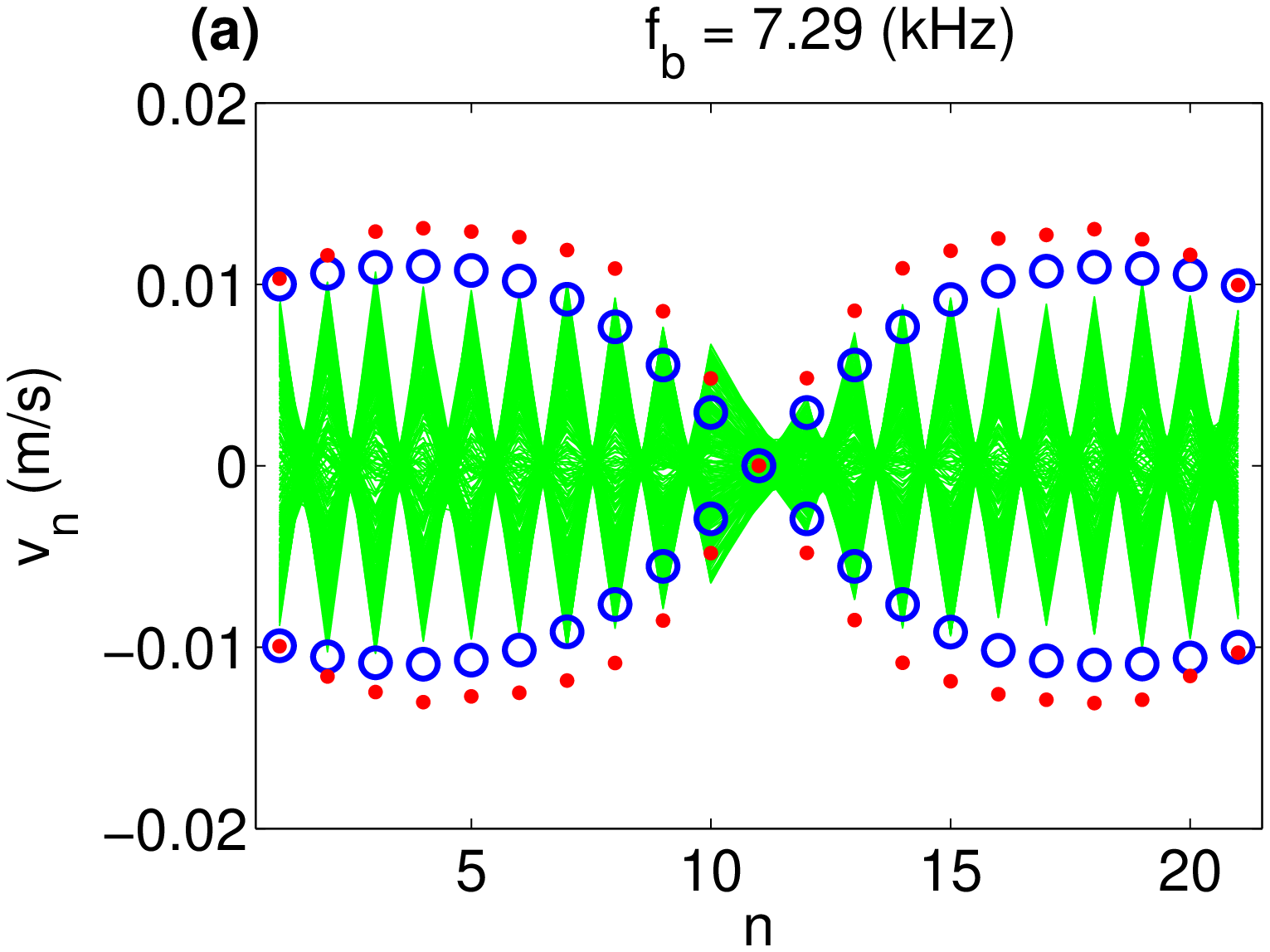,width=.33 \textwidth, height=4cm}
\epsfig{file=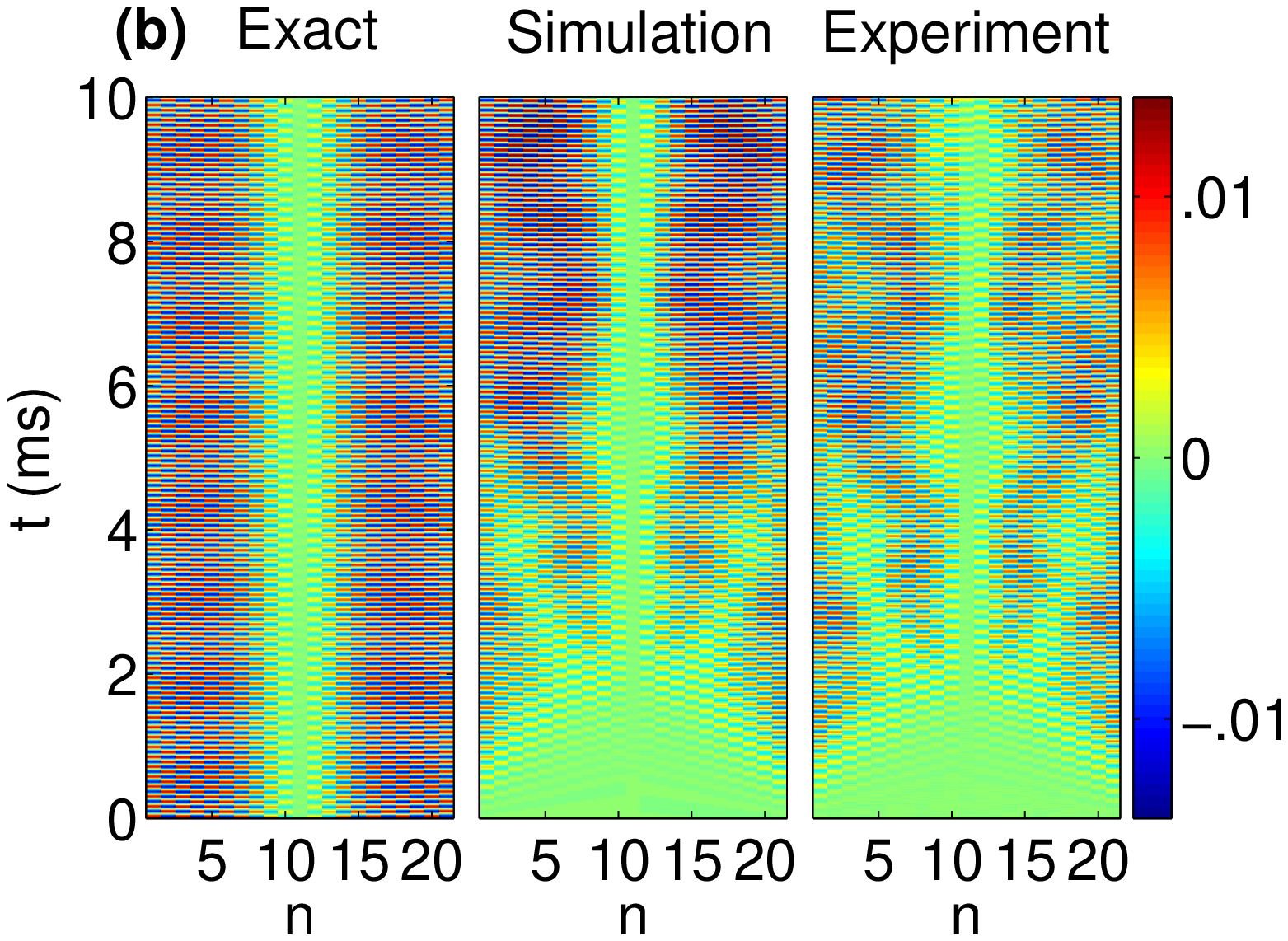,width=.33 \textwidth, height=4cm}
\epsfig{file=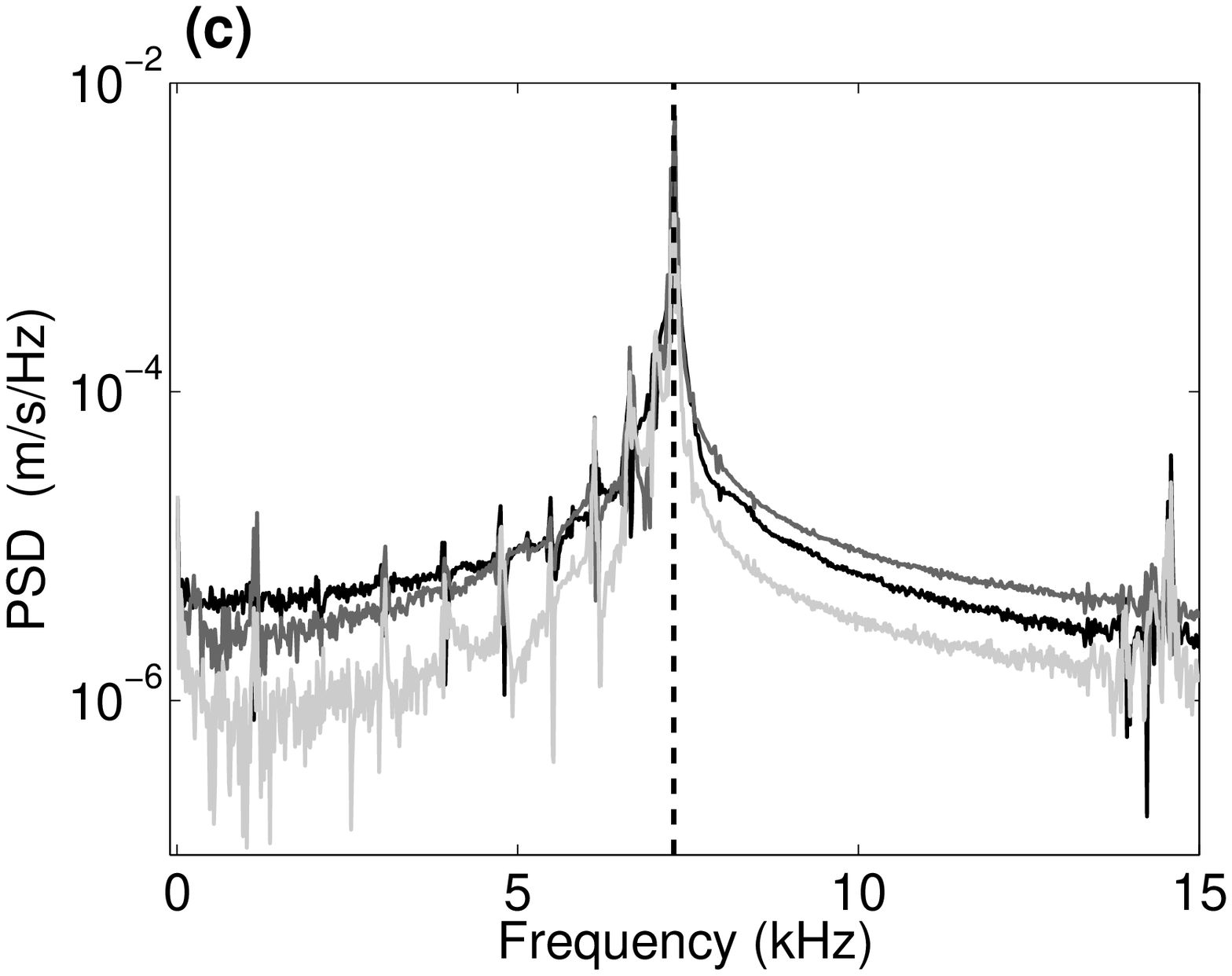,width=.33 \textwidth, height=4cm}
}
\caption{(color online) \textbf{(a)}  Experimentally measured velocities versus bead number for $f_b = 7.29$ kHz
and $a=0.198 \mu$m. The entire time series of each bead location is shown as a superposition (green lines). Extrema as predicted by the simulation for the first $10$ ms (red points) and numerically exact dark breather (blue circles) is also shown. \textbf{(b)}
Space-time contour plots of the exact breather, a transiently simulated 
one from zero initial data, and experimental evolution, 
also from zero initial data, leading to the same state. 
Color intensity corresponds to velocity (m/s). 
\textbf{(c)} Experimentally measured power spectral density (m/s/Hz) of bead 1 (darkest line), bead 5 (dark line) and bead 10 (lightest line) for an actuation amplitude of $a=0.198$ $\mu$m. The vertical dashed line corresponds to the driving frequency $f_b = 7.29$ kHz.}
\label{fig:profiles1}
\end{figure*}

Granular crystals, which consist of closely packed arrays of particles~\cite{nesterenko1,sen08} that interact elastically, 
are relevant for numerous applications such as shock and energy absorbing 
layers~\cite{dar06,hong05,fernando,doney06}, 
actuating devices \cite{dev08}, acoustic lenses \cite{Spadoni}, acoustic diodes \cite{Nature11}, sound scramblers \cite{dar05,dar05b}
and energy harvesters \cite{Nature11}.  
At a fundamental level, granular crystals have been
shown to support defect modes~\cite{Theo2009},
bright discrete 
breathers in dimer chains (i.e., bearing 
two alternating masses)~\cite{Theo2010} and surface variants 
thereof~\cite{hooge12}.
Very recently, dark breathers were theoretically
proposed in a Hamiltonian variant of the system as the 
sole discrete breather configuration that can arise in a ``monoatomic''
chain~\cite{dark}.

Our aim here is to interweave these
coherent structures (dark discrete breathers) with such a
lattice setting of interest to applications (granular crystals).
In particular, we intend to combine: theoretical considerations
through a realistic damped-driven model of a monoatomic granular crystal,
numerical computations identifying the canonical solutions
of this system (namely dark breathers and most importantly
{\it multibreather} generalizations thereof) and  
physical experiments that reveal their existence and robustness.


{\it Theoretical \& Experimental Setup.} A dark breather (DB) is a time-periodic structure with tails that are oscillating at a finite amplitude (as opposed
to bright breathers where the oscillation amplitude asymptotes
to $0$ as the lattice index ${n \rightarrow \infty}$);
see e.g.  Fig.~\ref{fig:profiles1}. 
For example, the function 
$ (-1)^n  \alpha \tanh(\beta n) \cos (2 \pi f_b t)$, 
with $\alpha, \beta$ constants, 
can be thought of as  a dark breather with frequency $f_b$. 
Dark breathers were shown to have this form in several nonlinear lattice models including the Klein-Gordon lattice \cite{Alvarez02}, the Fermi-Pasta-Ulam lattice \cite{James03,Rey04}, and the (Hamiltonian) monoatomic granular crystal lattice \cite{dark}. Following the theoretical proposal of~\cite{dark}, we intend to use
a {\it destructive interference} mechanism to {\it spontaneously generate}
the DBs. We will 
 actuate the granular crystal at both of its boundaries with frequency $f_b$,
i.e. the frequency of our intended DB.  In order to induce a vanishing amplitude at the central site,
(and the density dip associated with a DB)
there should be an odd number of beads and out of phase actuation such that incoming waves 
from each boundary actuation will ``cancel each other out" at the center. 

\begin{figure}[t]
\centerline{
\epsfig{file=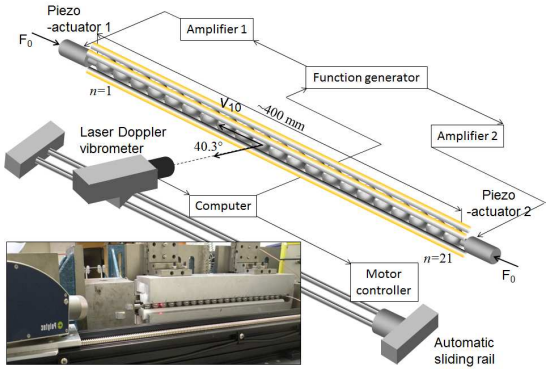,width=.45 \textwidth}
} 
\caption{(color online) Schematic of the experimental setup. The inset shows a digital image of the setup.}
\label{fig:exp}
\end{figure}

Figure~\ref{fig:exp} shows the schematic of the experimental setup consisting of a 21-sphere granular chain and a laser Doppler vibrometer (LDV, Polytec, OFV-534). The spheres have a radius $R=9.53$ mm and are made of chrome steel (with Young's modulus $E = 200$~GPa, Poisson ratio $\nu = 0.3$ and mass $M = 28.2$ g) \cite{exp1}. They are supported by four polytetrafluoroethylene rods allowing free axial vibrations of the particles, while restricting their lateral motions. Both ends of the granular chain are compressed with a static force ($F_0 = 10$~N, as in \cite{exp2}), and they are driven by two piezoelectric actuators that are powered individually by an external function generator and two amplifiers. The LDV measures the individual particles' velocity profiles and produces a full-field map of the granular chain dynamics. 

To model the experimental setup
we incorporate to the
standard granular crystal model of~\cite{nesterenko1}
a simple description of the dissipation~\cite{Nature11} 
and 
out-of-phase actuators on the left and right boundaries:
\begin{equation} \label{eq:gc}
\begin{array}{l}
M \ddot{u}_n = A [ \delta_0 +   u_{n-1} -   u_{n}]_+^{3/2}  \\ \hspace{.7cm} - A [ \delta_0 +   u_{n} - u_{n+1}]_+^{3/2}  -  \frac{M}{\tau} \dot{u}_n,  \\  
 u_0 = a \cos( 2 \pi f_b t), \, u_{N+1} = - a \cos( 2 \pi f_b t)
\end{array} 
\end{equation}
where  $N$ (odd) is the number of beads in the chain, $u_n(t)$ is the displacement of the $n$-th bead from the equilibrium position at time $t$,
$ A =  \frac{ E \sqrt{2R} }{ 3(1-\nu^2)} $,
$M$ is the bead mass and $\delta_0$ is an
equilibrium displacement induced by a static load $F_0=A 
\delta_0^{3/2}$. The bracket is defined by $[x]_+ = \mathrm{max}(0,x)$.
The strength of the dissipation is captured by the parameter $\tau$,
whereas $a$  and $f_b$ represent the amplitude and frequency of the
actuation, respectively.   
In what follows, we fix $\tau=0.5$ {ms} based
on experimental observation and treat $a$ and $f_b$ as the sole control parameters.

The pass band of the linearized equations of motion is  $[0, f_0 ]$  where 
$f_0 =   \sqrt{  \frac{3 A }{2 M \pi^2} }  \delta_0^{1/4} $     
is the cutoff frequency. For the parameter values used herein 
$f_0 = 7.373$ kHz. In addition to this band, the boundary actuators drive 
additional modes with frequency
$f_b$ (which for $f_b \notin [0,f_0]$ will correspond to 
surface modes, explained briefly below).  
To obtain the dark breathers experimentally (and in numerical simulations)
we use the following excitation procedure:   
the frequency of actuation is chosen within the pass band  $f_b\in[0,f_0]$
with zero initial conditions.
In this case, the propagation of plane waves and their
subsequent destructive interference {\it spontaneously} produces
the dark breathers.  
In order to ensure the robust
formation of a DB and avoid the onset of transient, large amplitude traveling waves \cite{nesterenko1}, we tune
the actuation amplitude in the simulations and experiments to be
increased linearly from zero to the desired amplitude $a$ over some fixed 
number of  periods
(we chose eight).

Dark breathers obtained with this excitation procedure are compared with
numerically {\it exact} (up to a prescribed tolerance) 
periodic solutions of~\eqref{eq:gc} 
by computing roots of the
map $u(T_b) - u(0)$ using e.g. a Newton method \cite{Flach2007}.  
Note that the breather frequency and actuation frequency are both $f_b=1/T_b$ by construction. 

\begin{figure}[t] 
\centerline{
\epsfig{file=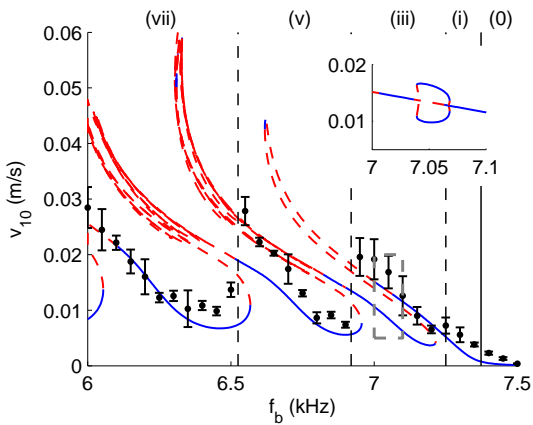,width=.45 \textwidth}
}
\caption{(color online) Bifurcation diagram illustrating the various
dark breather/multibreather branches. 
Specifically, the maximum velocity of bead 10 versus frequency $f_b$ is shown 
with $a=0.198$ $\mu$m. The smooth curves correspond to the numerically exact dark breathers (solid blue lines correspond to linearly stable solutions and red-dashed lines are unstable regions). 
The linear cutoff frequency $f_0 = 7.373$ kHz  is indicated by the solid 
vertical black  line. The black markers represent the experimentally measured
mean values with standard deviations given by the error bars, which were obtained using four experimental runs.  
The inset in the top-right is a zoom of the gray dashed box showing a representative example of additional branches that emerge from (and disappear 
back into) the main branch.}
\label{fig:freq_bif}
\end{figure}

{\it Results.} Figure~\ref{fig:profiles1} (a-b) 
shows a numerical and experimental dark breather excitation for the
``typical'' values of $f_b=7.29$ kHz and
$a=0.198$ $\mu$m, and the corresponding numerically exact solution at those parameter values.
The strong structural similarity in space to an exact dark breather (see e.g. panel (a)) suggests that both are near the steady-state after $10$ ms.  It is then not surprising that the motion of the beads is periodic,
as shown by a power spectral density (PSD) plot (panel (c)). 
The detailed comparison of the space-time
evolution of the numerically exact solution, simulation, and experiment in panel (b)
shows that after the initial
transient stage of the dynamics, a DB is formed.
We note that the maximum strain $|u_n- u_{n+1}|$ of 
the solution is about $60\%$ of the precompression, confirming that the structures reported here
are a result of the nonlinearity of the system.

\begin{figure*}
\centerline{
\epsfig{file=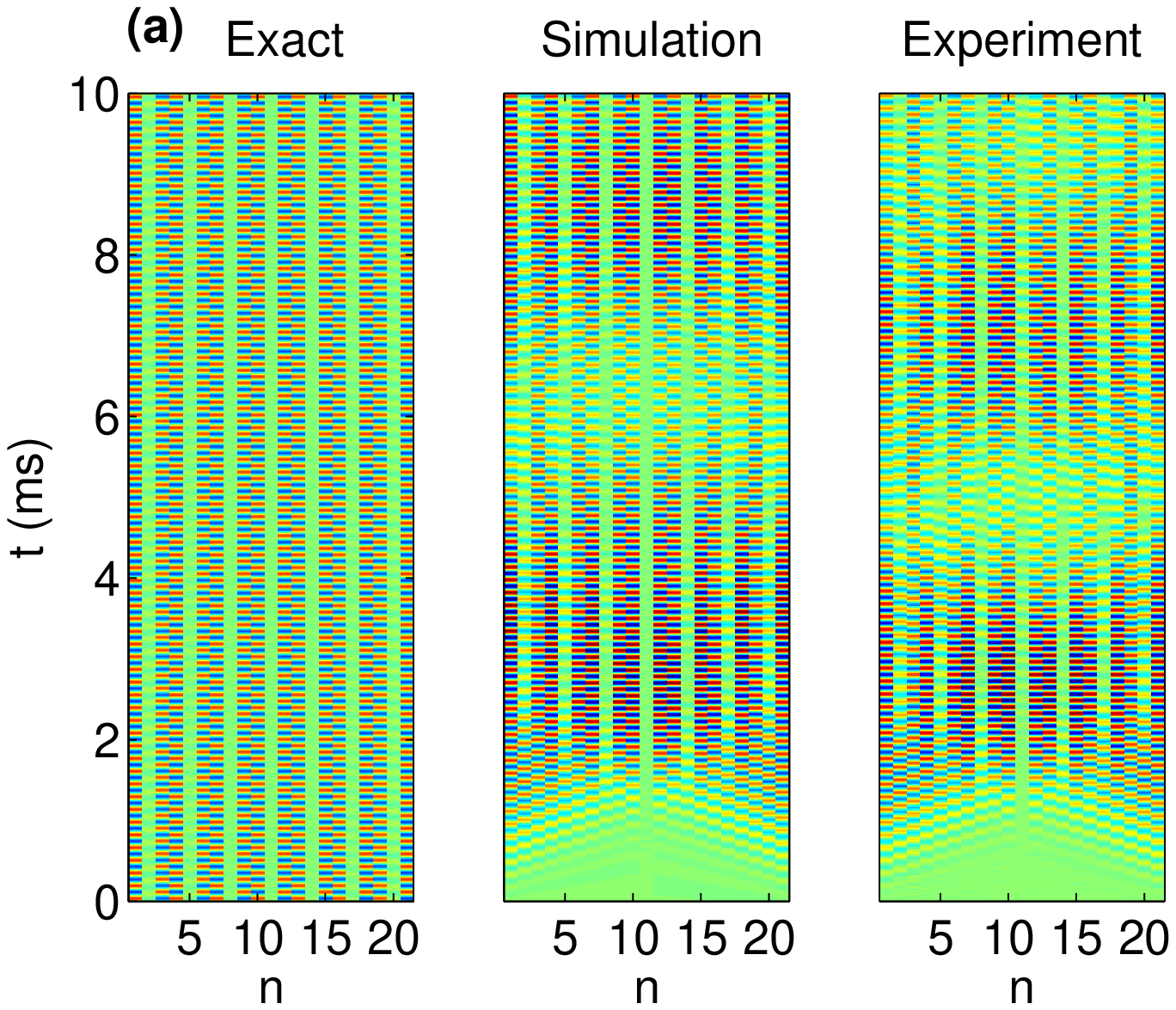,width=.33 \textwidth}
\epsfig{file=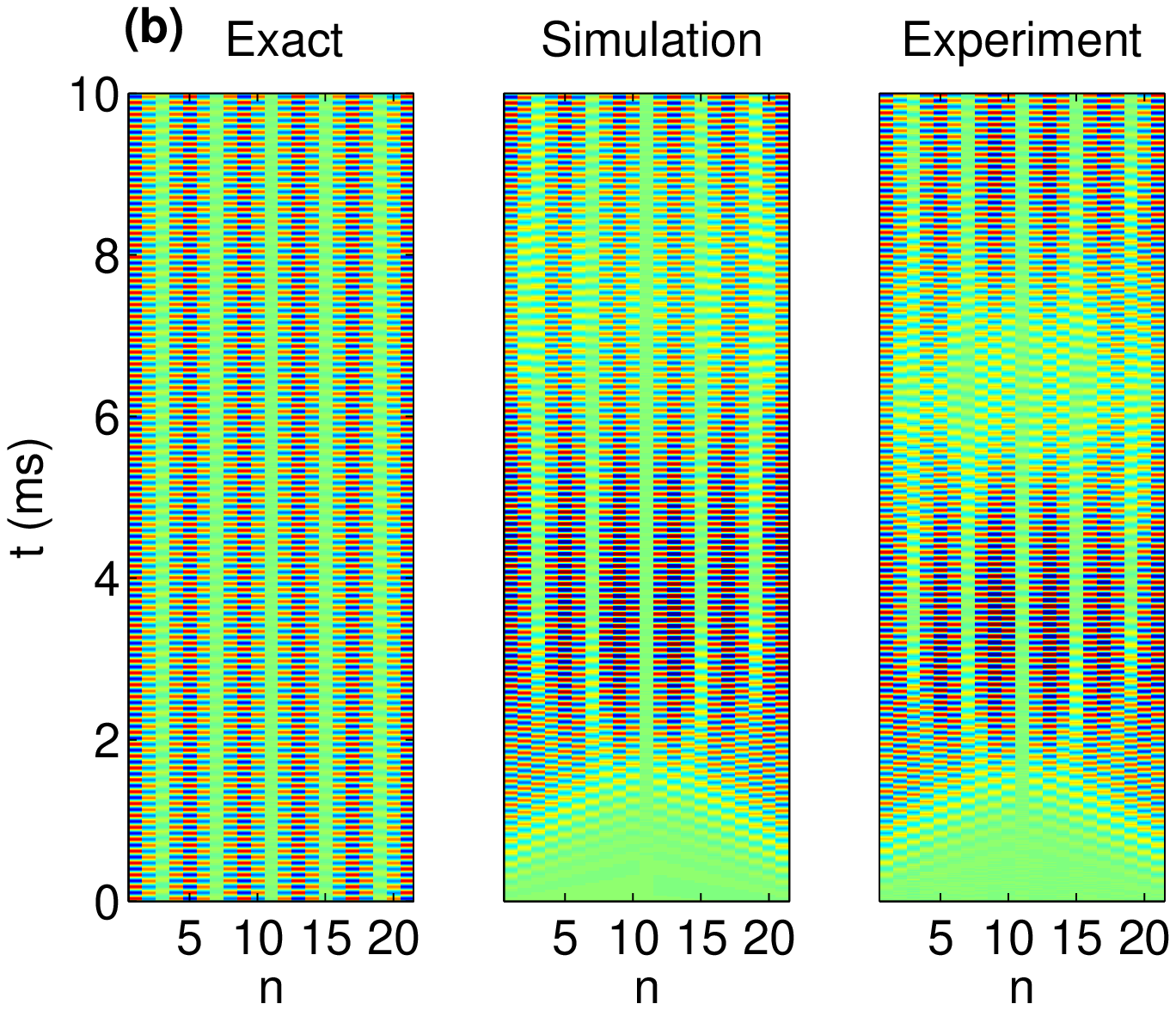,width=.33 \textwidth}
 \epsfig{file=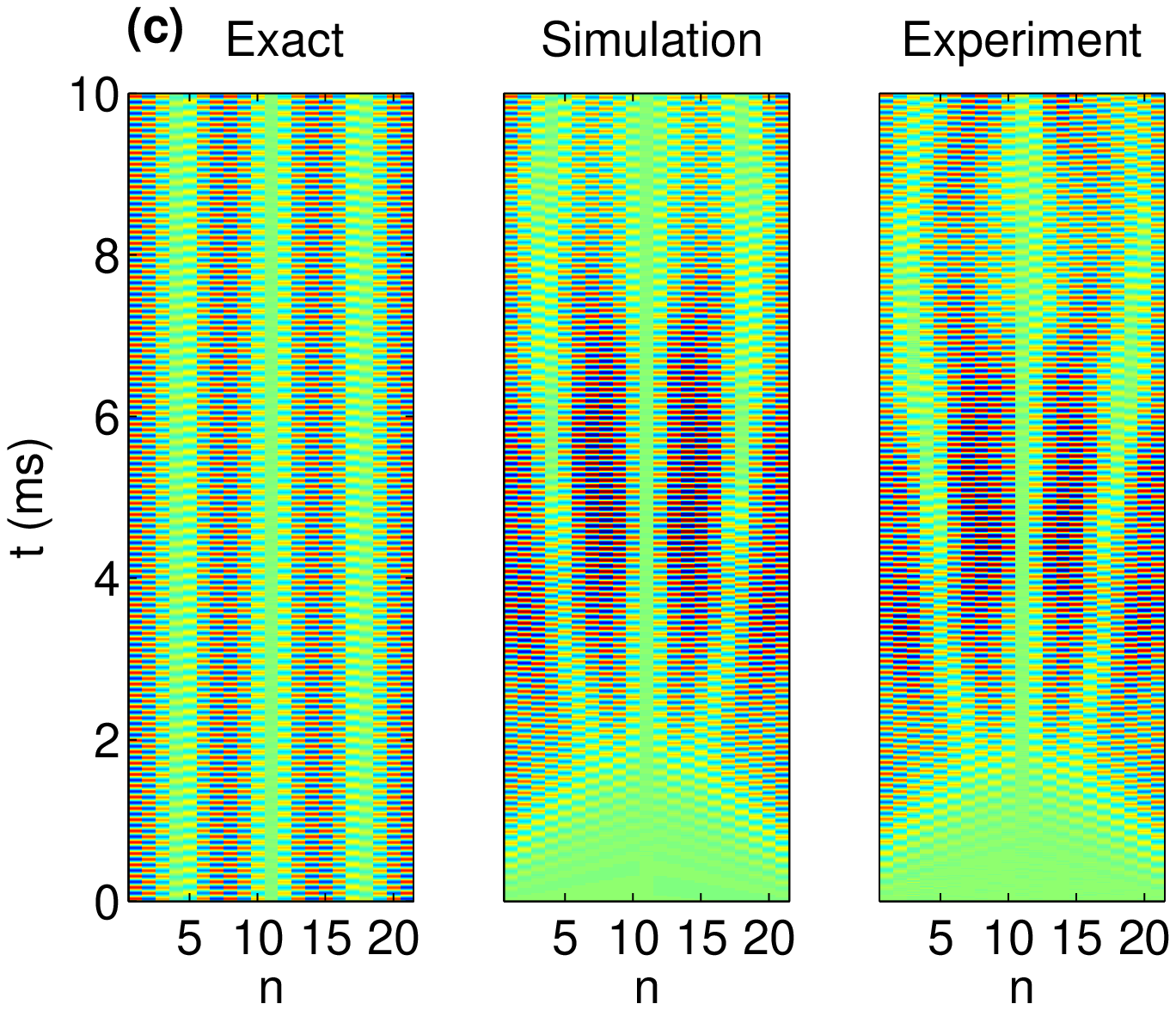,width=.33 \textwidth} 
}
\caption{(color online) Space-time contour plots of the numerically exact, simulated, and experimental evolution leading to a dark breather
with $a = 0.198$ $\mu$m. 
\textbf{(a)} 7~dip solution with $f_b = 6.35$ kHz. 
\textbf{(b)} 5~dip solution with  $f_b = 6.8$ kHz.
\textbf{(c)} 3~dip solution with $f_b = 7.15$ kHz.
Color intensity corresponds to velocity (m/s) where the color legend of each panel is the same as in Fig.~\ref{fig:profiles1}. }
\label{fig:contours}
\end{figure*}

We 
now study the full bifurcation diagram of the dark breathers
shown in Fig.~\ref{fig:freq_bif} and the corresponding profiles/evolutions
illustrated in Fig.~\ref{fig:contours}.
A numerical continuation reveals that, for a fixed driving amplitude  (here of
$a=0.198$ $\mu$m), the dark breathers
and multibreathers appear to be located on a \emph{single} coiling solution branch. Each fold represents the collision
of two breather families, such as a dark breather with a multibreather. Due to the ``coiling'' structure and the nature of
excitation simulations and experiments considered herein (which start with a zero initial state and
converge to the state with the lowest (non-zero) energy)
the transition from one multibreather family to the next as the forcing 
frequency is increased (or decreased) is not smooth:   
saddle node bifurcations cause the solution to ``jump'' from lobe to lobe. For example,
%
%
%
%
in Fig.~\ref{fig:freq_bif} the regions
labeled (vii), (v), (iii) and (i) correspond, respectively, 
to the number of density dips seen in the (experimental) space-time contour
plot. The label (0) corresponds to the surface discrete
(bright) breathers localized near the boundaries rather than at
the center of the domain.  
The experimentally measured solutions are indicated by black markers with error bars in Fig.~\ref{fig:freq_bif}.  Within each respective region, the 
observed dark breathers 
correspond to the branch with lowest energy.
For example, in region (vii)
at $f_b = 6.35$ kHz a solution with 7 dips emerges from the interference introduced at the boundaries
(see Fig.~\ref{fig:contours}a).
However, as we gradually increase the frequency, 
the outermost dips approach the boundaries until the 
solution collides and vanishes
in a saddle-node bifurcation with an intermediate 7 dip solution, i.e. the 
second lowest branch shown in region (vii) of Fig.~\ref{fig:freq_bif} (see the supplemental material examples of intermediate solution profiles).  
As a result of the disappearance of this branch, the lowest
branch in region (v) is made up of solutions with only 5 dips, e.g. for $f_b = 6.80$ kHz (see Fig.~\ref{fig:contours}b). The cascade of saddle-node bifurcations continues, 
as we gradually increase the frequency.
A solution with only 3 dips emerges in region (iii),
 e.g for $f_b = 7.15$ kHz (see Fig.~\ref{fig:contours}c). 
Finally, a single dip dark breather emerges for frequencies in region (i) e.g. for $f_b=7.29$ kHz (see Fig.~\ref{fig:profiles1}).
Thus, we conclude that the actuation frequency 
that is chosen will {\it dictate}
(``dial in'') the number of dips that will emerge
upon actuation.

In addition to the saddle-node events connecting the
observable to the intermediate branches in this coiled structure
connecting each pair of branches to the next, there 
also exist symmetry-breaking pitchfork bifurcations (see e.g. the inset of Fig.~\ref{fig:freq_bif} and the supplemental material).

Unlike the Hamiltonian model of \cite{dark},
the solutions do not have vanishing amplitude as the cut off frequency is 
approached (since the actuation amplitude remains fixed),
and thus dark breathers with frequencies above the cut off exist as well. 
However, the dip
is much broader within the gap, and corresponds rather to a surface
breather mode~\cite{hooge12}.  

\begin{figure*}
 \centerline{
\epsfig{file=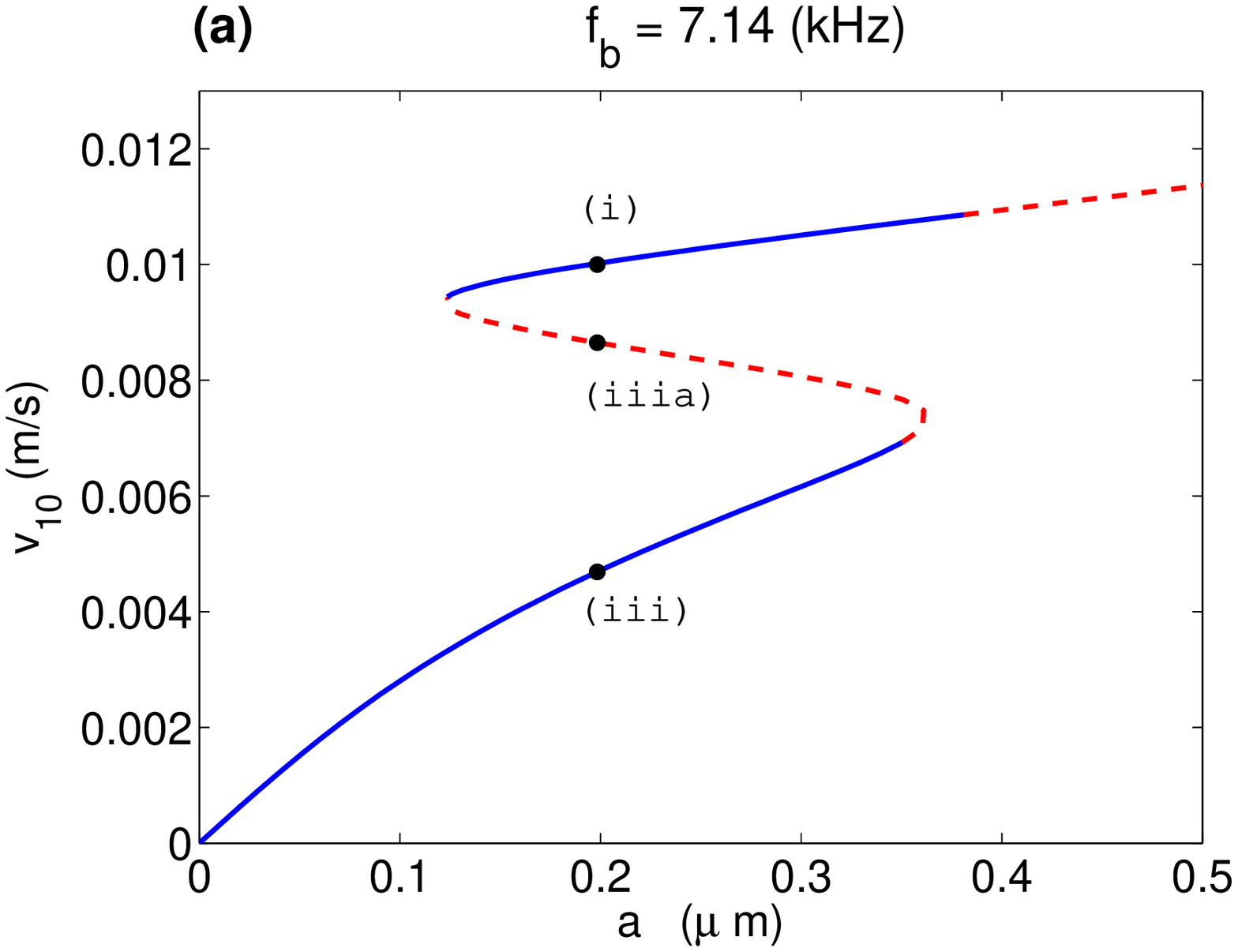,width=.33 \textwidth, height=4cm}
\epsfig{file=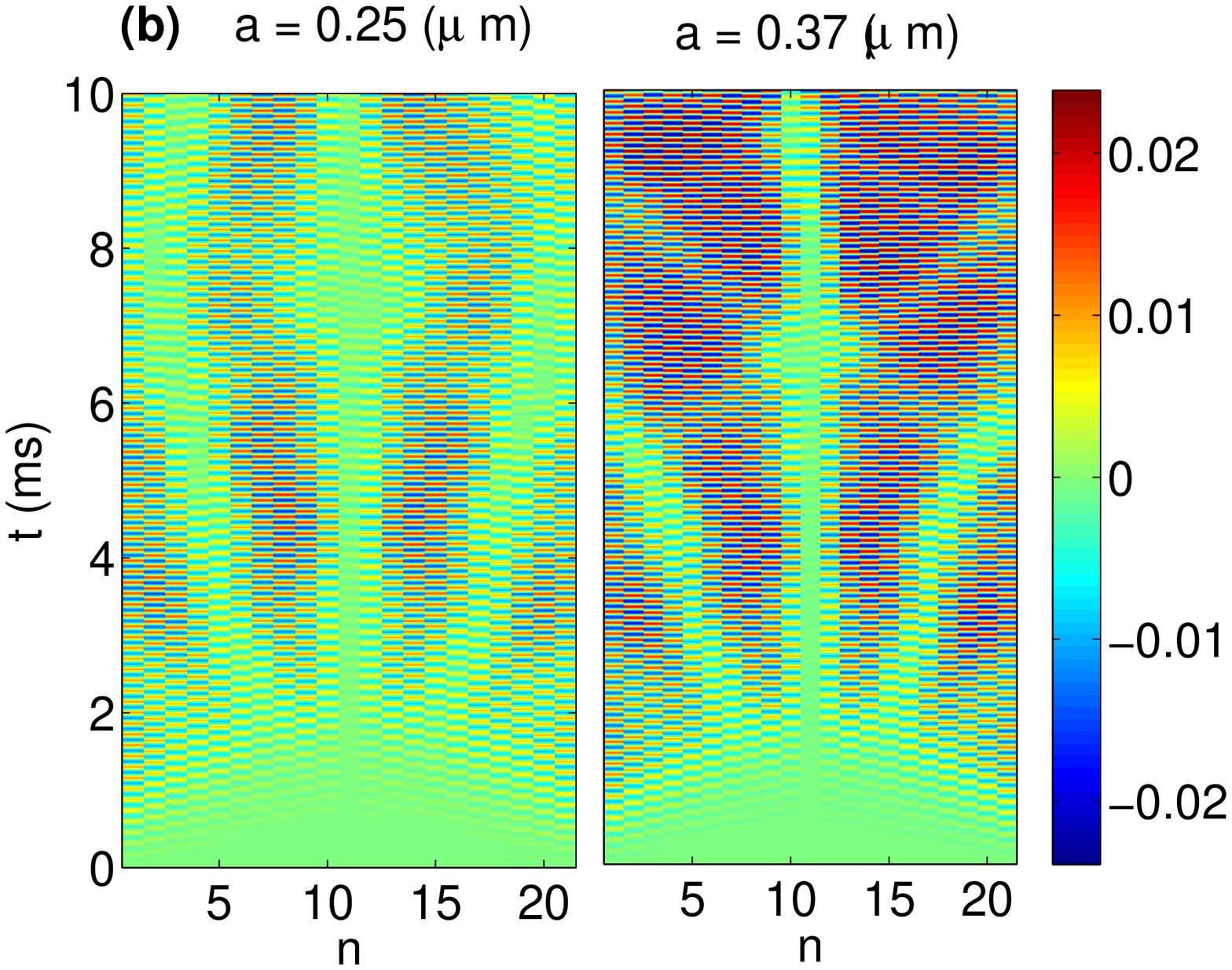,width=.33 \textwidth, height=4cm}
\epsfig{file=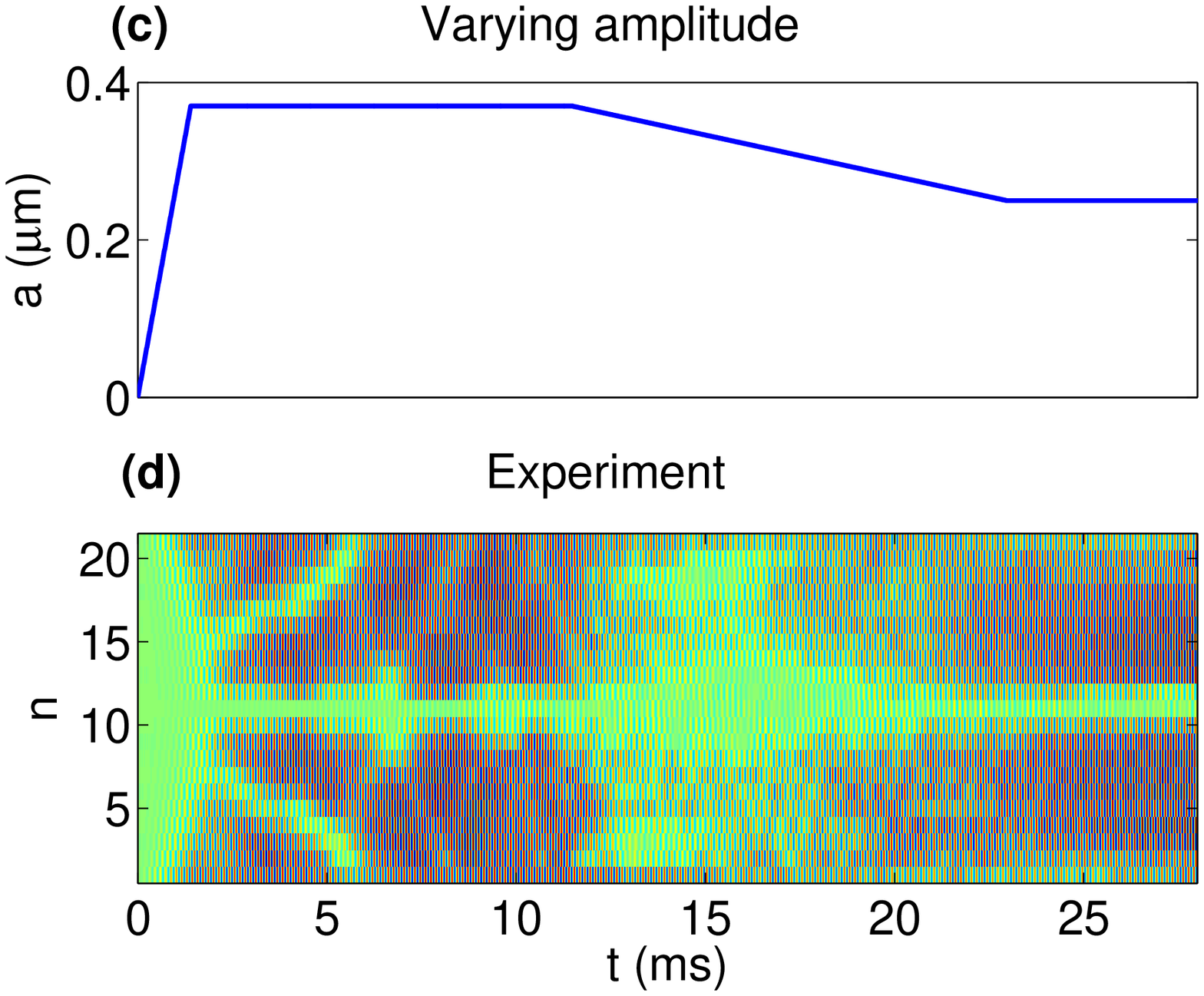,width=.33 \textwidth, height=4cm}
}
\caption{(color online) \textbf{(a)} Maximum velocity of bead 10 of the exact dark breather versus actuation amplitude $a$ with the frequency fixed as $f_b = 7.14$ kHz.  The black points indicate the value $a=0.198$ $\mu$m which corresponds to the frequency continuation shown in Fig.~\ref{fig:freq_bif}.   
One can
see that the single dip breather (i) and 3 dip breather (iii) are connected by an intermediate unstable solution (iiia). \textbf{(b)} Experimental space-time contour plots of the velocity. By choosing the actuation amplitude $a$ appropriately
one can excite a 3 dip breather ($a=0.25$ $\mu$m in the left contour plot) or a single dip breather ($a=0.37$ $\mu$m in the right contour plot). Note that, the 3 dip breather terminates 
at $a \approx 0.36$ $\mu$m in the panel (a). \textbf{(c)}  The  actuation amplitude profile used to excite a high energy dark breather.  After the initial linear ramping, the amplitude is kept at $a=0.37$ $\mu$m for about $10$ ms, 
which will excite
a single dip breather. The single dip breather is still maintained even when decreasing the amplitude to $a=0.25$ $\mu$m. \textbf{(d)} Space-time contour plot corresponding to the actuation amplitude profile shown in (c). Color intensity corresponds to velocity (m/s), where the color bar is the same as 
in (b). Note that, if one actuates a resting chain with an actuation amplitude of $a=0.25$ $\mu$m then a three dip breather will emerge 
(see i.e. the left panel of  (b))  thus revealing the system's bistability.
}
\label{fig:amp_bif} 
\end{figure*}

To investigate the dynamical stability of the obtained states, 
a Floquet analysis was carried out to compute the multipliers associated
with the DBs (details given in supplemental material). 
In Fig.~\ref{fig:freq_bif} a red-dashed line is shown if
a purely real instability is present whereas a blue solid line is shown otherwise.
An oscillatory instability, which can also be present here, is a result of a pair
of Floquet multipliers crossing the unit circle outside of the real line. 
This indicates the existence of Hopf (Neimark-Sacker) bifurcation points and hence of possibly nearby 
quasi-periodic solutions and solutions with period $m T_b$, where $m$ 
is an integer. The investigation
of such states, however, is outside the scope of this letter.

In addition to the multibreather control via frequency, 
one can also control the number of dips of the dark breather by varying the amplitude. 
Consider for example the three solution branches shown in region (iii) of Fig.~\ref{fig:freq_bif} at $f_b = 7.14$ kHz.
At this frequency the system is bistable, with the three dip and single
dip breather being stable and an intermediate 3 dip solution being unstable.  These three solutions
were continued in amplitude (see Fig.~\ref{fig:amp_bif}a)  revealing a canonical hysteresis loop between the stable
3 dip and single dip solution branches. This enables even 
experimentally a jump between the different solution types 
(as shown in Fig.~\ref{fig:amp_bif}b).   
One can excite the single dip breather in the parameter region $a \in (0.12, 0.37)$ by first driving to the single dip breather with $a>0.37$, and then
adjusting the amplitude to some value in the region $a \in (0.12, 0.37)$, see Fig.~\ref{fig:amp_bif}(c-d) for example.
Note that, exciting a resting chain for a fixed frequency in that region will yield the three dip breather, see e.g. the left panel of Fig.~\ref{fig:amp_bif}b,
a feature indicative of the system's bistability.

{\it Conclusions/Future Challenges.} The damped-driven granular crystal system
has been shown to provide 
access to a rich family of dark breather 
and multibreather solutions. The system possesses an intricate bifurcation 
diagram where the stable single- and multi-dip solutions are interlaced
via unstable intermediate branches. The diagram contains a large number
of saddle-node bifurcations and associated fold points, as well
as pitchfork symmetry-breaking points. This structure provides not only 
hysteresis loops and 
multi-stability regimes, but also a remarkable tunability. 
The
selection of driving frequency and amplitude 
enables a selection of multibreather
configurations 
robustly sustained by the dynamics. 
The coherent structures produced herein
could be utilized towards controllable energy funneling and harvesting
within granular media.

{\it Acknowledgements.}
The authors would like to thank G. Theocharis for useful discussions. 
Support from US-NSF (CMMI 844540, 1200319 and 1000337) and US-AFOSR 
(FA9550-12-1-0332)
is appreciated.


\clearpage

\begin{minipage}[t]{.9 \textwidth}
\large \bf
\begin{center} SUPPLEMENTAL MATERIAL:\\
Damped-Driven Granular Crystals: An Ideal Playground for Dark Breathers and Multibreathers
\end{center}
\vspace{1.8cm}
\end{minipage}

\paragraph{Parameter values and scaling.}
The material/geometric parameter $A$ is given by 
$$A    =  \frac{ E \sqrt{2R} }{ 3(1-\nu^2)}, $$
where the bead radius and mass are $R=9.53$ mm and $M=28.2$ (g) respectively,
the Young's modulus is $E = 200\cdot10^9$ GPa, and
the Poisson ratio is $\nu=0.3$. The applied precompression force
was $F_0 = 10$ N. Nondimensionalization of the equations of motion reveals that 
the solutions live in a three-dimensional parameter space
(e.g., $a,f_b,$ and $\tau$). However, we use $\tau$ as a fixed parameter  ($\tau=0.5$ ms throughout the text)
and allow $a$ and $f_b$ to vary.


\begin{figure}[b]
 \epsfig{file=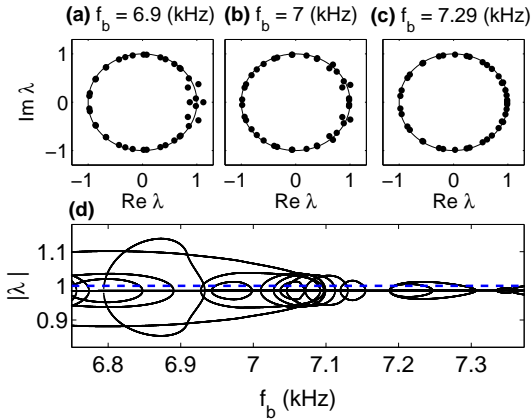,width=.47 \textwidth} 
\caption{(color online) Floquet multipliers of the single dip dark breather for $a=0.198$ $\mu$m.  \textbf{(a)}  At $f_b = 6.90$ kHz the solution is unstable due to oscillatory instabilities and Floquet multipliers on the real line.
\textbf{(b)} The Floquet  multipliers on the real line have retreated within the unit circle at  $f_b = 7.00$ kHz but the oscillatory instabilities remain.  \textbf{(c)} Finally, at $f_b = 7.29$ kHz all instabilities have vanished, and the solution is asymptotically stable. \textbf{(d)} Magnitude
of Floquet multipliers for frequencies (kHz) in the interval $f_b \in (6.75,7.4)$. A blue dashed line at $|\lambda| =1$ is also shown to help identify regions of
asymptotic stability.  }
\label{fig:FM}
\end{figure}

\paragraph{Details of the Newton-Raphson and the linear stability computations.} The numerically obtained dark
breathers of period $T_b=1/f_b$ reported in the text were found by computing roots of
the map ${F = u(T_b) - u(0)}$, where $u(0)$ is the initial state vector of the (numerically realized) stroboscopic map
and $u(T_b)$ is the solution of the equations of motion at time 

\phantom{ aaaaaaaaaaaaaaaaaaaaaaaaaaaaaaaaaaaaa} \vspace{2.9cm}

\noindent $T_b$.
The Jacobian of  $F$, which is used in the Newton iterations, is of the form
$V(T_b)-I$, where $I$ is the appropriate identity matrix, $V$
is the solution to the variational equation $V' = DF V$ with initial 
condition $V(0) = I,$ and  $DF$ is the Jacobian of the equations of 
motion evaluated at $u$. 
%
The Floquet multipliers for a solution were  obtained by computing the eigenvalues of
the monodromy matrix, which is $V(T_b)$ upon convergence of the Newton-Raphson scheme.
In both the main text and  supplement, we focus on instabilities associated with saddle-node bifurcations and pitchfork bifurcations;  therefore, we are 
chiefly interested in the Floquet multipliers on the  (positive) real line.
However, there can 
also be  oscillatory instabilities, which correspond to  complex-conjugate pairs of  Floquet multipliers  lying outside of the unit circle.
Figure~\ref{fig:FM} shows examples of the Floquet multipliers for three single-dip solutions. The first two examples, in Fig.~\ref{fig:FM}(a-b), are unstable, but the third example, in Fig.~\ref{fig:FM}c,  is asymptotically stable.
A plot of the magnitude of the multipliers for fixed $a=0.198$ $\mu$m and various $f_b$ is shown in Fig.~\ref{fig:FM}d.

A complication revealed by the Floquet analysis is that even though asymptotically stable solutions are possible, the damping is weak enough that these solutions may not be  realizable in the 10 ms window considered experimentally. For example, for $f_b = 7.29$ kHz and $a=0.198$ $\mu$m 
there is a dark breather with 
Floquet multipliers
in the interval  $| \lambda | \in [0.974,0.998]$, and thus it is asymptotically stable. 
Because these multipliers are so near the unit circle, converging to a fixed point by repeatedly applying $F $ to some initial condition, which is analogous to what happens when the experiment is run for multiple  periods, requires a large number of iterations  or  running an experiment for a long time. Therefore, due to the 10 ms window used, we are often only able to see the experiment \textit{approach} the dark breather and do not see  it \textit{fully converge} to the dark breather.

Conversely, there also exist 
oscillatory instabilities 
on the solution branch for some parameter intervals.
%
Although the dark breather is now unstable, we do not expect an oscillatory instability to manifest in the 10 ms window if we start with an initial condition near enough to the unstable dark breather.   
In that case, we have found that
the effects of the oscillatory instability are typically not observable 
in computations until around $t=300$ ms. 
The reason for this is that the magnitudes of the Floquet multiplier pairs associated with the oscillatory instabilities are often closer to unity than   the magnitudes of unstable Floquet multipliers on the real line; this is why we indicate only the instabilities due to Floquet multipliers on the real line in Fig.~3 even though oscillatory instabilities may also be present.

\begin{figure}
\centerline{
\epsfig{file=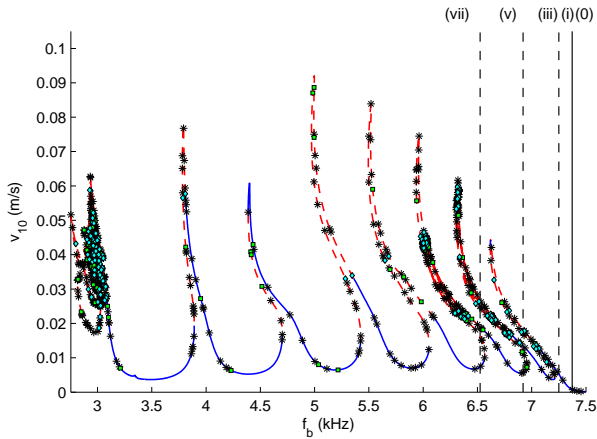, width=0.5\textwidth}}
\caption{(color online) The full branch of solutions shown in Fig.~3. 
Blue regions of the branch have no unstable eigenvalues on the positive real line, but they may have oscillatory instabilities. 
The symbols indicate the presence of a bifurcation: black stars are Neimark-Sacker bifurcations to $T^2$ tori, green squares are period-doubling bifurcations, and cyan diamonds are pitchfork bifurcations.   
The dashed lines indicate the regions where the $N$-bump solutions appear as shown in Fig.~3.
}
\label{fig:full_bifur}
\end{figure}

\begin{figure}
 \centerline{ 
  \epsfig{file=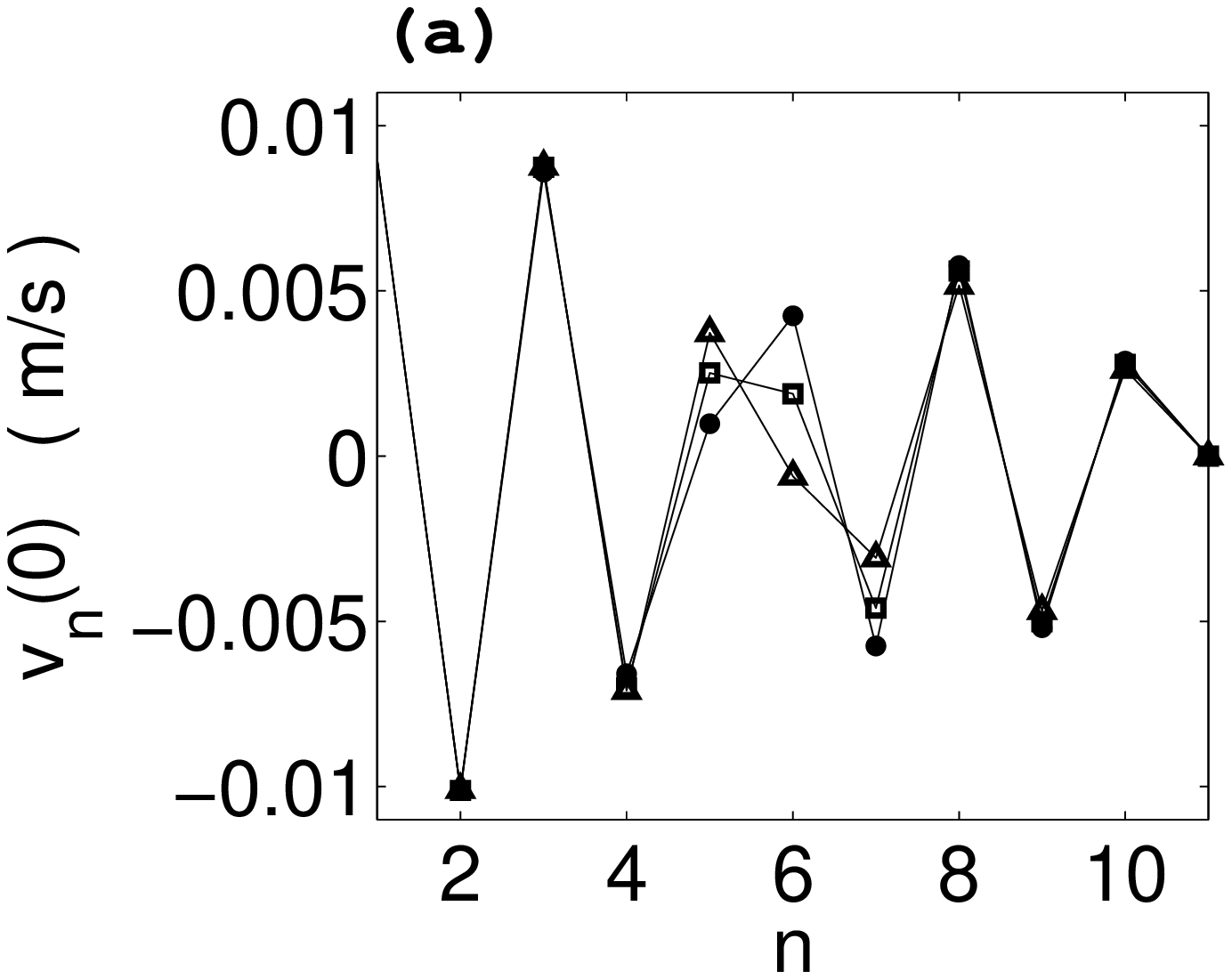,width= .25 \textwidth}
 \epsfig{file=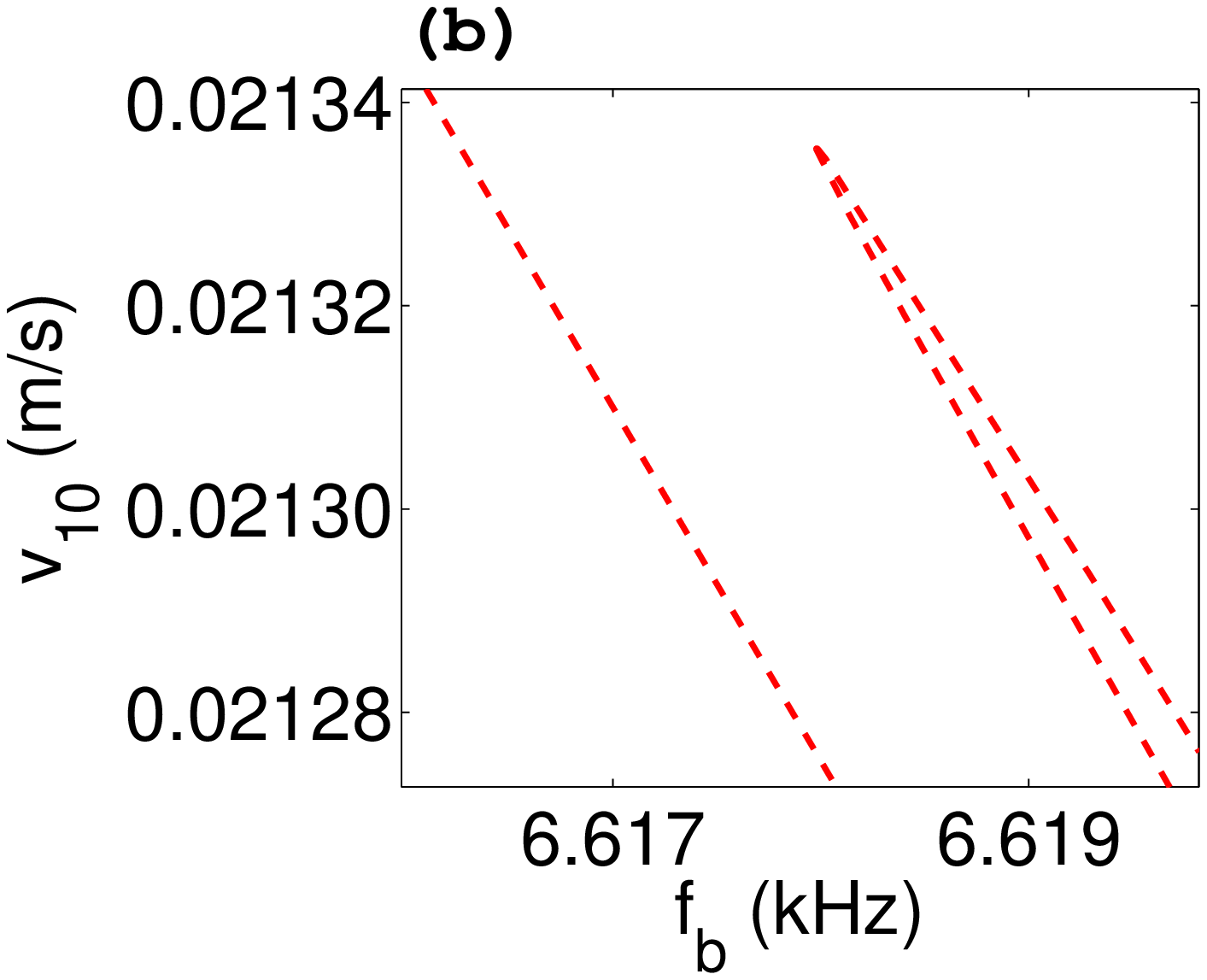,width=.25 \textwidth} }
\caption{(color online) \textbf{(a)} Velocity profile for three variants of a 3-dip breather for
$a=0.198 \mu m$ and $f_b = 6.62$ kHz indicated by square, circle, and triangle
markers at $t=0$. Since the solutions have odd symmetry with respect to the center bead ($n=11$), only the left part of the chain is shown. \textbf{(b)} Tight
zoom of Fig.~3 of
the manuscript where three distinct (but connected) branches can be seen (which cannot be
discerned in the former figure). These three branches correspond to the three profiles on the
left.}
\label{fig:zoom}
\end{figure}

\paragraph{Bifurcations and secondary branches of periodic solutions.}
The branch of solutions shown in Fig.~3 possesses an intricate bifurcation structure, composed of several coiling branches, and generates many secondary and tertiary branches of periodic solutions. 
Figure~\ref{fig:full_bifur} shows a ``zoomed out'' version of this 
coiled branch and indicates the bifurcations that occur on it.  The black stars, green squares, and cyan diamonds indicate the presence of Neimark-Sacker, period-doubling, and pitchfork bifurcations respectively; the latter two are the points on the main branch where the secondary branches of periodic solutions are born.
Here, we computed the branch of multibreather solutions below the 6 kHz frequency level that was studied experimentally, and found that the same structure continues down to the 3 kHz level where the ``top'' of the branch becomes significantly more complicated.
%
%
A ``zoomed in'' version of one of the coils, shown in Fig.~\ref{fig:zoom} reveals that the branch actually coils several times in tight regions
in parameter space, which cannot be seen in Fig.~3 of the main text.
%
%
  
\begin{figure*}
 \centerline{ 
 \epsfig{file=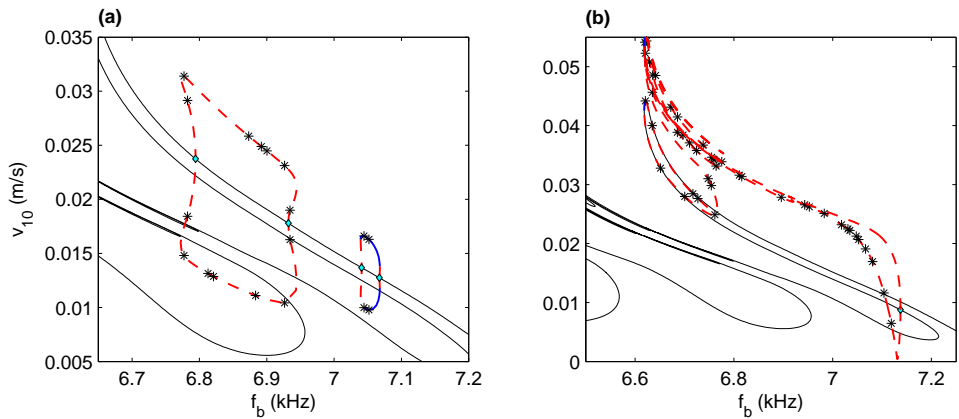,width= .8\textwidth}  
 }
\caption{(color online) Examples of secondary branches with torus bifurcations indicated with black stars and pitchfork bifurcations indicated by cyan diamonds.
As in Fig.~3, the blue solid regions are where all the Floquet multipliers on the real line are in or on the unit circle,  and the red, dashed regions of the curve are where they are not.  The main solution branch is indicated in black. 
 {\bf (a)} Plot of a pair of pitchfork loops, each of which   consists of at least two subcritical pitchfork bifurcations and two pairs of saddle-node bifurcations.
We have also observed that some pitchfork loops  have intervals \textit{without} non-oscillatory instabilities.
 {\bf (b)}  Plot of  pitchfork branches  that is not part of a pitchfork loop.
These secondary branches can have regions without non-oscillatory instabilities, as demonstrated by the blue ``specks'' at the top of the plot, but in our experience, these regions are short lived.
   }
\label{fig:secondary}
\end{figure*}

The secondary branches of interest in this letter are initiated by pitchfork bifurcations, either due to a regular pitchfork bifurcation or  as part of a  pair in what we call a   pitchfork loop.
To illustrate the difference, we present a prototypical example for each type in Fig.~\ref{fig:secondary}.

\begin{figure}
\centering
\epsfig{file=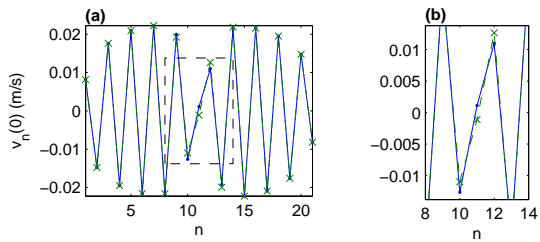, width=0.45\textwidth}
\caption{(color online) \textbf{(a)} Plot of the velocity profiles at $t=0$ on either ``arm'' of the pitchfork loop shown in the inset of Fig.~3 that demonstrate the symmetry breaking that occurs when the forcing amplitude is at a maximum.
The blue, solid line is the ``lower'' branch while the green, dashed line is the upper branch in the figure.
\textbf{(b)} Zoom of (a) around beads 8 --12.
}
\label{fig:loop_symmetry}
\end{figure}

The prototypical pitchfork loop, shown in  Fig.~\ref{fig:secondary}a, consists of a pair of subcritical pitchfork bifurcations and at least two pairs of saddle-node bifurcations that are on a single, secondary branch of solutions.
These pitchfork loops appear to be relatively short lived, i.e., the pitchfork bifurcation that ``opens'' the loop is near (in terms of arclength) to the pitchfork bifurcation that ``closes'' the loop.
As a result, although the main branch may have an unstable Floquet multiplier on the real line, there always appears to be a pair of ``nearby'' solutions that are qualitatively similar to the main branch with the addition of a small component that breaks the symmetry that solutions have on the main branch. However, the two secondary branches that comprise the pitchfork loop appear to be symmetric to each other; starting from a pitchfork bifurcation, both branches appear to have the same sequence of bifurcations, and the only difference between solutions on the branches is the direction in which the symmetry-breaking component manifests itself.
 An example of this is shown for a pair of saddle-node bifurcation points in Fig.~\ref{fig:loop_symmetry}. 
The velocity profile, which is an odd function  on the main branch if the center bead is taken as the origin, loses this symmetry on either branch of the pitchfork loop.
This should be contrasted with the ``coiled'' solutions in Fig.~\ref{fig:zoom} that are perturbations of one another, but {\em still have velocity profiles that are odd functions,}  again using the center bead as the origin.
Furthermore, this symmetry appears to hold at all times and not only at the phase where the stroboscopic map was used.
Lastly, as demonstrated in Fig.~\ref{fig:secondary}a,  pitchfork loops were also observed on segments of the branch that already have an unstable, real Floquet multiplier.

There are also regular pitchfork bifurcations that produce solution branches that do not quickly ``close'' and are not symmetric with respect to each other. 
We refer to the secondary branches created by these bifurcations as pitchfork branches, and give an example of one in  Fig.~\ref{fig:secondary}b.
Unlike  pitchfork loops, these secondary branches only seem to appear on the parts of the branch where stability has already been lost, usually through a saddle-node bifurcation, and appear
to be  unstable for the values of $f_b$ and $v_{10}$ that were studied experimentally. However, we have found that they can regain stability for very small regions at  large values of $v_{10}$. Computationally, we have observed multiple Neimark-Sacker bifurcations (indicated by black stars) and period doubling bifurcations on these  branches, but surprisingly, we have not observed additional   pitchfork bifurcations.

Although these types of secondary branches are quite complex and exist for non-negligible intervals of $f_b$, they may be difficult to observe experimentally as they are unstable in the regions of experimental interest.

%
%
%
%

Due to the number of bifurcations that generate secondary branches and the complexity and additional bifurcations that appear on the secondary solution branches, this system apparently gives rise to a veritable ``zoo'' of dynamics and displays a vast array of different nonlinear behaviors.
To complicate matters further, there are also additional branches of periodic solutions with longer periods  (e.g., period-two solutions of the map $F$ that have period $2T_b$) whose shapes and bifurcations are also non-trivial.
Both these larger period solutions and solutions on the pitchfork branches generate tertiary branches, which  can also be stable for the frequencies of interest here.
As shown in Fig.~\ref{fig:full_bifur} and Fig.~\ref{fig:secondary}, the main branch and the secondary branches are filled with Neimark-Sacker bifurcations and there are very few intervals on the branch are not unstable in an oscillatory fashion.  
As a result, the solutions that are observed are  likely to lie on  stable, slowly-modulating tori, and thus may  truly be quasi-periodic rather than periodic. 

Overall, the bifurcation diagram in Fig.~3  imparts a good qualitative understanding of the dynamics that appear experimentally.
While there are numerous details and additional subtleties to this system, which are interesting and will be the focus of future theoretical work, 
they do not appear to appreciably impact the behavior of the system  
over the 10 ms windows considered experimentally.


\begin{thebibliography}{10}

\bibitem{Flach2007} S. Flach and A.~V. Gorbach, Phys. Rep. {\bf 467}, 1 (2008).

\bibitem{moti} 
F. Lederer, 
G.I. Stegeman, D.N. Christodoulides, G. Assanto,
M. Segev, and Y. Silberberg,
Phys. Rep. {\bf 463}, 1 (2008).

\bibitem{sievers}
M. Sato, 
B.E. Hubbard, and A.J. Sievers,
Rev. Mod. Phys.  {\bf 78}, 137 (2006)

\bibitem{alex} 
P. Binder, 
D. Abraimov, A.V. Ustinov, S. Flach, and Y. Zolotaryuk,
Phys. Rev. Lett. {\bf 84}, 745 (2000); 
\\
E. Tr{\'{\i}}as, 
J.J.  Mazo, and T.P. Orlando,
Phys. Rev. Lett. {\bf 84}, 741 (2000).


\bibitem{lars3}
L.Q. English, 
M. Sato, and A.J. Sievers,
Phys. Rev. B {\bf 67}, 024403 (2003); 
\\
U.T. Schwarz, 
L.Q. English, and A.J. Sievers,
Phys. Rev. Lett. {\bf 83}, 223 (1999).

\bibitem{swanson}
B.I. Swanson, 
J.A. Brozik, S.P. Love, G.F. Strouse, A.P. Shreve,
A.R. Bishop, W.-Z. Wang, and M.I. Salkola, 
Phys. Rev. Lett. {\bf 82}, 3288 (1999).

\bibitem{Peybi} 
M. Peyrard, Nonlinearity {\bf 17}, R1 (2004).

\bibitem{Morsch} O. Morsch and M. Oberthaler, Rev. Mod. Phys. {\bf 78}, 179
(2006).

\bibitem{Chabchoub} A. Chabchoub, O. Kimmoun, H. Branger, N. Hoffmann, D. Proment, M. Onorato, and N. Akhmediev
Phys. Rev. Lett. \textbf{110} 124101 (2013) 

\bibitem{weller} A. Weller, J. P. Ronzheimer, C. Gross, J. Esteve, M. K. Oberthaler, D. J. Frantzeskakis, G. Theocharis, and P. G. Kevrekidis
Phys. Rev. Lett. {\bf 101}, 130401 (2008); S. Stellmer, C. Becker, 
P. Soltan-Panahi, E.-M. Richter, S. D{\"o}rscher, M. Baumert, J. Kronj{\"a}ger, 
K. Bongs, and K. Sengstock
Phys. Rev. Lett. {\bf 101}, 120406 (2008).

\bibitem{djf} D.J. Frantzeskakis,
J. Phys. A {\bf 43}, 213001 (2010).

\bibitem{carr} W. Tong, M. Wu, L.D. Carr, and B.A. Kalinikos
Phys. Rev. Lett. {\bf 104}, 037207 (2010).

\bibitem{kanshu} A. Kanshu, C. R{\"u}ter, D. Kip, J. Cuevas,
P.G. Kevrekidis, Eur. Phys. J. D {\bf 66}, 182 (2012).

\bibitem{nesterenko1} V.~F. Nesterenko, {\it Dynamics of Heterogeneous Materials} (Springer-Verlag, New York, NY, 2001).

\bibitem{sen08} S. Sen, J. Hong, J. Bang, E. Avalos, and R. Doney, Phys. Rep. {\bf 462}, 21 (2008).


\bibitem{dar06} C. Daraio, V.~F. Nesterenko, E.~B. Herbold, and S. Jin, Phys. Rev. Lett. {\bf 96}, 058002 (2006).

\bibitem{hong05} J. Hong, Phys. Rev. Lett. {\bf 94}, 108001 (2005).

\bibitem{fernando} F. Fraternali, M.~A. Porter, and C. Daraio, Mech. Adv. Mat. Struct. {\bf 17}(1), 1 (2010).


\bibitem{doney06} R. Doney and S. Sen, Phys. Rev. Lett. {\bf 97}, 155502 (2006).

\bibitem{dev08} D. Khatri, C. Daraio, and P. Rizzo, SPIE {\bf 6934}, 69340U (2008).



\bibitem{Spadoni} A. Spadoni and C. Daraio, Proc Natl Acad Sci USA, {\bf 107}, 7230, (2010).


\bibitem{Nature11} N. Boechler, G. Theocharis, C. Daraio,  Nature Materials \textbf{10}, 665 (2011). 


\bibitem{dar05} C. Daraio, V.~F. Nesterenko, E.~B. Herbold, and S. Jin, Phys. Rev. E {\bf 72}, 016603 (2005).

\bibitem{dar05b} V.~F. Nesterenko, C. Daraio, E.~B. Herbold, and S. Jin, Phys. Rev. Lett. {\bf 95}, 158702 (2005).


\bibitem{Theo2009} G. Theocharis, M. Kavousanakis, P.~G. Kevrekidis, C. Daraio, M.~A. Porter, and I.~G. Kevrekidis, Phys. Rev. E {\bf 80}, 066601 (2009);
S. Job, F. Santibanez, F. Tapia and F. Melo, Phys. Rev. E {\bf 80}, 
025602 (2009);
Y. Man, N. Boechler, G. Theocharis, P. G. Kevrekidis, and C. Daraio,
Phys. Rev. E {\bf 85}, 037601 (2012).


\bibitem{Theo2010} N. Boechler, G. Theocharis, S. Job, P.~G. Kevrekidis, M.~A. Porter and C. Daraio, Phys. Rev. Lett. {\bf 104}, 244302 (2010);
G. Theocharis, N. Boechler, P.~G. Kevrekidis,  S. Job, M.~A. Porter, and C. Daraio, Phys. Rev. E {\bf 82}, 056604 (2010). 

\bibitem{hooge12} C. Hoogeboom, Y. Man, N. Boechler, G. Theocharis, P. G. Kevrekidis, 
I. G. Kevrekidis and C. Daraio, Euro. Phys. Lett. {\bf 101 }, 44003 (2013) 

\bibitem{dark} C. Chong, P.G. Kevrekidis, G. Theocharis, and C. Daraio 
Phys. Rev. E. \textbf{87}  042202 (2013)

\bibitem{Alvarez02} A.  Alvarez, J.F.R Archilla, J. Cuevas, and F. R. Romero,
New J. of Phys. {\bf 4},  72  (2002)

\bibitem{James03} G. James, J. Nonlinear Sci. \textbf{13}, 27 (2003).

\bibitem{Rey04}  B. Sanchez-Rey, G. James, J. Cuevas, and J.~F.~R. Archilla, Phys. Rev. B {\bf 70}, 014301 (2004).




\bibitem{exp1} http://www.efunda.com

\bibitem{exp2} F. Li, L. Yu, and J. Yang. J. of Phys. D: Appl. Phys.  \textbf{46}, 155106 (2013)
























\end{thebibliography}
\end{document}